# Ultra directive antenna via transformation optics

## P. H. Tichit, S. N. Burokur, A. de Lustrac

Institut d'Electronique Fondamentale,
Centre Scientifique d'Orsay 91400 Orsay , France
Fax: + 33169154090; e-mail: paul-henri.tichit@u-psud.fr

**Abstract**

Spatial coordinate transformation is used as a reliable tool to control electromagnetic fields. In this paper, permeability and permittivity tensors of a metamaterial able to transform an isotropically radiating source to a compact ultra-directive antenna in the microwave domain are calculated. The directivity of this antenna is competitive with regard to conventional directive antennas (horn, reflector antennas) even with its smaller dimensions. Numerical simulations using Finite Element Method (FEM) are performed to illustrate these properties. A reduction of the electromagnetic material parameters is also proposed for an easy fabrication of this antenna from existing materials. Following that, the design of the proposed antenna using a layered metamaterial is presented with simple metal-dielectric structures.

## 1. Transformation formulation

We consider here a line source radiating in a cylindrical vacuum space. In order to produce a highly directive emission, a physical space where lines θ = constant become horizontal is generated.

Mathematically the transformation used can be expressed as:

$$\begin{cases} x' = \dfrac{2L}{d}\sqrt{x^2+y^2} \\ y' = \dfrac{e}{\pi}\arctan(\dfrac{y}{x}) \\ z' = z \end{cases} \quad \text{With} \quad -\dfrac{\pi}{2} \leq \arctan(\dfrac{y}{x}) \leq \dfrac{\pi}{2} \quad (1)$$

where x', y' and z' are the coordinates in the transformed rectangular space and x, y, z are those in the initial cylindrical space. In the cylinder we assume free space, with isotropic permeability and permittivity tensors $\varepsilon_0$ and $\mu_0$ and the following transformations are used to obtain the material parameters of the new rectangular space:

$$\varepsilon^{i'j'} = \dfrac{J_i^{i'}J_j^{j'}\varepsilon_0\delta^{ij}}{\det(J)} \quad \text{and} \quad \mu^{i'j'} = \dfrac{J_i^{i'}J_j^{j'}\mu_0\delta^{ij}}{\det(J)} \quad \text{with} \quad J_\alpha^{\alpha'} = \dfrac{\partial x'^\alpha}{\partial x^\alpha} \quad (2)$$

where $J_\alpha^{\alpha'}$ and $\delta^{ij}$ are respectively the Jacobian transformation matrix of the transformation of (1) and the Kronecker symbol.

By substituting the new coordinate system in the tensor components, and after some simplifications, the following material parameters are derived:

$$\overline{\overline{\varepsilon}} = \begin{pmatrix} \varepsilon_{xx}(x',y') & 0 & 0 \\ 0 & \varepsilon_{yy}(x',y') & 0 \\ 0 & 0 & \varepsilon_{zz}(x',y') \end{pmatrix}\varepsilon_0 \qquad \overline{\overline{\mu}} = \begin{pmatrix} \mu_{xx}(x',y') & 0 & 0 \\ 0 & \mu_{yy}(x',y') & 0 \\ 0 & 0 & \mu_{zz}(x',y') \end{pmatrix}\mu_0 \quad (3)$$

where

$$\varepsilon_{xx}(x',y') = \mu_{xx}(x',y') = \frac{\pi}{e}x' \quad \varepsilon_{yy}(x',y') = \mu_{yy}(x',y') = \frac{1}{\varepsilon_{xx}(x',y')} \quad \varepsilon_{zz} = \mu_{zz} = \frac{d^2\pi}{4eL^2}x' \quad (4)$$

The divergence of $\varepsilon_{yy}$ near x' = 0 creates an "electromagnetic wall" with $\varepsilon_{yy} \to \infty$ at the left side of the rectangular area. This left side also corresponds to the radiating source transformed from the center line source of the cylindrical space. We can note the simplicity of the $\varepsilon_{xx}$ and $\varepsilon_{zz}$ which present a linear variation.

## 2. Simulations results

In this section, FEM based numerical simulations with Comsol Multiphysics [9] are used to design and characterize this transformed directive antenna. As the line source of the half right cylindrical space becomes a radiating plane in the transformed rectangular space, an excitation is inserted at the left side of the structure. This space is delimited by metallic boundaries on the upper and lower sides and at the left side of the rectangular space representing the metamaterial having dimensions 15 cm × 15 cm. Three operating frequency have been considered here 5, 10 and 40 GHz, corresponding respectively to $e/\lambda$ = 2.5, 5 and 20. Then for a half-power beamwidth of 13.5° at 10 GHz we obtain a directivity of 23.6 dB, implying a ratio $e/\lambda$ = 5. This directivity is comparable with that of a parabolic reflector antenna of the same size and is greater than that of a wideband [2 - 18 GHz] dual polarized FLANN horn antenna where the directivity varies from 10 dB to 23 dB.

The directivity increases strongly when we move towards higher frequencies. It goes from 23.6 dB at 5 GHz to 29,5 dB at 10 GHz and 42 dB at 40 GHz.

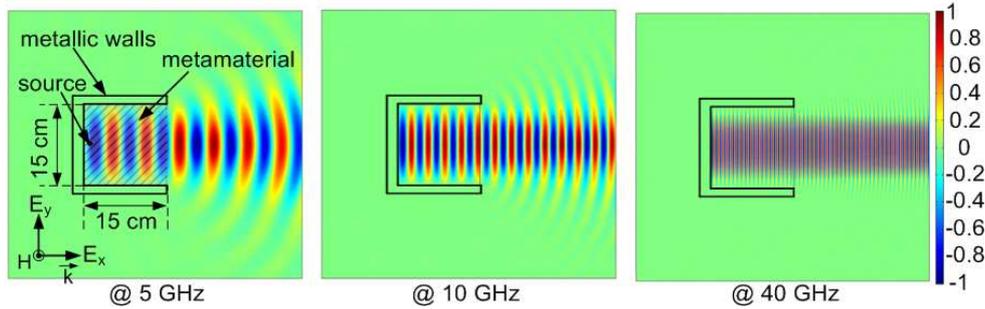

Fig. 1: Normalized Electric Field distribution in Transverse Electric polarization .

## 3. Layered metamaterial

The previous material has a continuous variation of permittivity along the x-axis (the propagation direction).We therefore adopt a method which consists in discretizing the permittivity profile in ten layers of different thickness and permittivity. We assume a constant variation $\Delta\varepsilon = \varepsilon_{max}/n$ at the end (rear interface) of each layer with $\varepsilon_{max} = \pi^2$ and n=10. Knowing that at the rear interface of layer $i$ $\varepsilon_{xx} = i\Delta\varepsilon$, we can therefore deduce the thickness $\Delta x_i$. Then for convenience, we apply to each different layer the value of $\varepsilon_{xx}$ corresponding to $x = x_i$ with $x_i$ lying at the middle of each layer. Noting $a = (\pi/e)^2$, the following relations are derived:

$$\Delta x_i = \sqrt{i\frac{\Delta\varepsilon}{a}} - \sqrt{(i-1)\frac{\Delta\varepsilon}{a}} \quad \text{with} \quad x_i = \frac{\sqrt{i\frac{\Delta\varepsilon}{a}} + \sqrt{(i-1)\frac{\Delta\varepsilon}{a}}}{2} \quad \text{for } 0 < i < n+1 \quad (5)$$

Simulations using Comsol Multiphysics are performed with the different thicknesses and permittivities determined with (5). A highly directive magnetic field distribution can be noted at 10 GHz for the TM wave incidence with the source embedded in the layered metamaterial. The normalized far field radiation pattern of the layered metamaterial-based antenna (dashed) agrees very well with the continuous metamaterial-based one (continuous), therefore indicating the easiness of fabrication.

## 4. Conclusion

A metamaterial structure is proposed in an antenna system to achieve the manipulation of directivity via spatial coordinate transformation. Numerical simulations are performed to show that the directivity of this metamaterial-based antenna is competitive with conventional directive antennas such as horn and reflector antennas. By embedding a radiating surface in the optimized metamaterial, a highly directive emission equivalent to parabolic antennas with similar dimensions is generated. The calculated reduced electromagnetic material parameters simplify the design procedure. For an easier fabrication with existing commercial dielectrics, a layered structure composed of homogeneous and uniaxial anisotropic metamaterials has been presented. In all the three cases (theoretical, reduced and layered metamterials) the directivity of the antenna remains consistent.